\documentclass[aps,pra,twocolumn,superscriptaddress]{revtex4-1}

\usepackage{graphicx}
\usepackage{hyperref}
\usepackage{amsmath}
\usepackage{amssymb}
\usepackage{epstopdf}
\usepackage{multirow}
\usepackage{threeparttable}
\usepackage{array}
\usepackage{booktabs}
\usepackage{xcolor}

\begin{document}

\title{Highly-sensitive mid-infrared upconversion detection\\ 
based on external-cavity pump enhancement}
\author{Xiaohan Liu}
\affiliation{State Key Laboratory of Precision Spectroscopy, East China Normal University, Shanghai 200062, China}

\author{Kun Huang}
\email{khuang@lps.ecnu.edu.cn}
\affiliation{State Key Laboratory of Precision Spectroscopy, East China Normal University, Shanghai 200062, China}
\affiliation{Chongqing Key Laboratory of Precision Optics, Chongqing Institute of East China Normal University, Chongqing 401121, China}
\affiliation{Collaborative Innovation Center of Extreme Optics, Shanxi University, Taiyuan, Shanxi 030006, China}

\author{Wen Zhang}
\affiliation{State Key Laboratory of Precision Spectroscopy, East China Normal University, Shanghai 200062, China}

\author{Ben Sun}
\affiliation{State Key Laboratory of Precision Spectroscopy, East China Normal University, Shanghai 200062, China}

\author{Jianan Fang}
\affiliation{State Key Laboratory of Precision Spectroscopy, East China Normal University, Shanghai 200062, China}

\author{Yan Liang}
\affiliation{School of Optical Electrical and Computer Engineering, University of Shanghai for Science and Technology, Shanghai 200093, China}

\author{Heping Zeng}
\affiliation{State Key Laboratory of Precision Spectroscopy, East China Normal University, Shanghai 200062, China}
\affiliation{Chongqing Key Laboratory of Precision Optics, Chongqing Institute of East China Normal University, Chongqing 401121, China}
\affiliation{Shanghai Research Center for Quantum Sciences, Shanghai 201315, China}
\affiliation{Chongqing Institute for Brain and Intelligence, Guangyang Bay Laboratory, Chongqing, 400064, China}

\begin{abstract}
Sensitive mid-infrared (MIR) detection is highly demanded in various applications, ranging from remote sensing, infrared surveillance, environmental monitoring to industrial inspection. Among others, upconversion infrared detectors have recently attracted increasing attention due to the advantageous features of high sensitivity, fast response, and room-temperature operation. However, it remains challenging to realize high-performance passive MIR sensing due to the stringent requirement of high-power continuous-wave pumping. Here, we propose and implement a high-efficiency and low-noise MIR upconversion detection system based on pumping enhancement via a low-loss optical cavity. Specifically, a single-longitudinal-mode pump at 1064 nm is significantly enhanced by a factor of 36, thus allowing for a peak conversion efficiency of up to 22\% at an intra-cavity average power of 55 W. The corresponding noise equivalent power is achieved as low as 0.3 fW/Hz$^{1/2}$, which indicates at least a ten-fold improvement over previous results. Notably, the involved single-frequency pumping would facilitate high-fidelity spectral mapping, which is particularly attractive for high-precision MIR upconversion spectroscopy in photon-starved scenarios.
\end{abstract}

\maketitle

\section{Introduction}
The mid-infrared (MIR) spectrum not only contains abundant fingerprinting vibrational transitions for many molecules, but also covers multiple windows with high transmittance in the Earth's atmosphere. Ultra-sensitive MIR detection is of great interest in scientific, industrial, and defense fields, such as trace analysis,  pollution monitoring, biomedical diagnosis, remote sensing, and free-space communications \cite{Vodopyanov2020Book}. Compared with visible or near-infrared bands, the MIR spectral region offers additional thermal information, which is particularly favorable for target recognition and object tracking in low-light conditions \cite{Razeghi2014RPP}. In these aforementioned applications, it is of great interest to improve the MIR detection sensitivity for addressing the needs in photon-scared scenarios, for instance, long-distance operation, covert surveillance, or phototoxicity-free observation \cite{Hadfield2023Optica, Russo2022Photonics}. However, commercial MIR photodetectors based on narrow-bandgap semiconductors like HgCdTe, InSb, or PbSe usually suffer from high dark noise and low response speed, which results in a typical noise equivalent power (NEP) at the level of pW/Hz$^{1/2}$ \cite{Razeghi2014RPP, Fang2020LPR}. Recently, superconducting nanowire single-photon detectors have demonstrated impressively broad spectral responses up to far-infrared wavelengths, albeit with stringent operation condition of cryogenic cooling \cite{Taylor2023Optica, Pan2022OE, Chen2021SB}. Nowadays, significant efforts have continuously been devoted to developing high-sensitivity MIR sensors at room temperature, especially by resorting to low-dimensional materials \cite{Wang2019Small, Wu2021NR, Xue2023LSA} or novel nanophotonic structures \cite{Xomalis2021Science, Chen2021Science}.

In this context, frequency upconversion detectors have emerged as an indirect yet effective approach to performing MIR detection and imaging \cite{Barh2019AOP}. In this approach, the infrared radiation is nonlinearly transferred to the visible or near-infrared regime, where more mature devices or techniques are available for high-performance optical manipulation and detection \cite{Huang2022NC, Zeng2023LPR}. To date, such infrared upconversion detection has widely been investigated in many areas, such as free-space quantum key distribution \cite{Liao2017NP}, real-time MIR optical coherence tomography \cite{Israelsen2019LSA}, or high-speed hyperspectral videography \cite{Junaid2019Optica, Zhao2023NC, Fang2024NC}. Generally, there are two categories for the upconversion detectors. One relies on the pulsed pumping, which favors high conversion efficiency and low noise due to intense peak power and narrow time window \cite{Wolf2017OEA, Donnell2019PR, Rehain2020NC, Huang2021PR, Fang2023LSA}. On one hand, the involved ultrafast optical excitation enables high-precision gating in time-resolved measurement \cite{Wolf2017OEA} and three-dimensional imaging \cite{Rehain2020NC, Fang2023LSA}. On the other hand, such an active detection fashion is usually limited to cooperative targets within a relatively short capture range. Additionally, the intrinsic broad spectrum for ultrashort pulses inevitably degrades the spectral resolution in upconversion infrared spectroscopy due to the convolution operation in the parametric wave-mixing process \cite{Zheng2023LPR, Sun2014LPR}.

Alternatively, a continuous-wave (CW) pumping scheme is employed to facilitate passive infrared sensing \cite{Demur2017OL, Li2024OLT, Ge2023PRA} and high-precision spectroscopy \cite{Neely2012OL}. As the most common setting for pump sources, single-pass configuration favors great simplicity for optical alignment, yet the involved average power is typically required to be over tens of watts to implement efficient nonlinear conversion \cite{Demur2017OL, Li2024OLT}. The use of a photonics waveguide is beneficial to reduce the required pump power \cite{Neely2012OL, Buchter2009OL}, but the single-spatial-mode confinement excludes the imaging capability. Another way to approach high pump intensity is to leverage the optical cavity enhancement technique, where the nonlinear crystal is placed within the resonating light field \cite{Witinski2009AO, Huang2019OE}. For instance, the nonlinear crystal can be directly inserted into a laser resonator to perform the so-called intra-cavity enhancement, which has been used to implement sensitive thermal imaging \cite{Dam2012NP, Huang2017LP} and photon-counting ranging \cite{Yue2022RS, Widarsson2022AO}. In this case, the wavelength of the involved laser diode for laser generation is usually close to the spectral band of the upconverted signal \cite{Dam2012NP}, which imposes an increasing difficulty for noise suppression through spectral filtering, and typically results in a NEP about tens of fW/Hz$^{1/2}$ \cite{Pedersen2019PTL}. To this end, an external-cavity enhancing scheme has been investigated to decouple the pump source and enhancing cavity, hence leading to an improved system flexibility \cite{Albota2004OL, Wolf2017OE}. Moreover, the pump source can benefit from outstanding properties of single-frequency fiber lasers, like compact layout, low noise, and narrow linewidth \cite{Fu2017JOSAB}, which offers great potential to implement high-fidelity MIR upconverter for high-resolution spectroscopy \cite{Neely2012OL} and coherent optical communication \cite{Buchter2009OL}. Recently, the external cavity enhancement has been adopted to demonstrate a broadband MIR upconversion spectrometer, albeit with a relatively high NEP of about few pW/Hz$^{1/2}$ \cite{Wolf2017OE}. Therefore, it is appealing to further manifest the full potential for high-sensitivity MIR sensing by combining high-performance silicon photon counters.

Here, we implement a highly-sensitive MIR upconversion detection system based on external-cavity pump enhancement, which leads to a NEP as low as 0.3 fW/Hz$^{1/2}$, about ten-fold better than previously reported values for CW-pumping MIR upconversion detectors. The optical cavity is constructed with a finesse of 109, which facilitates a power-enhancing factor up to 36. At an intracavity power of 55 W, the conversion efficiency is about 22\% due to optimized spatial-mode matching among the involved optical fields within the resonator. The technical complexity of the cavity locking is mitigated by using a digital locking unit based on a field programmable gate array (FPGA), which enables automatic stabilization functionality and long-term operation. Moreover, a multi-pixel photon counter (MPPC) is used to further extend the dynamic range of the MIR upconversion detector, which indicates at least a 15-dB improvement to that of a single-pixel detector. Additionally, the single-frequency pump favors high-fidelity spectral mapping, which holds great potential in sensitive and precise MIR spectroscopy.

\section{Experimental setup}
\begin{figure*}[t!]
\includegraphics[width=0.90 \textwidth]{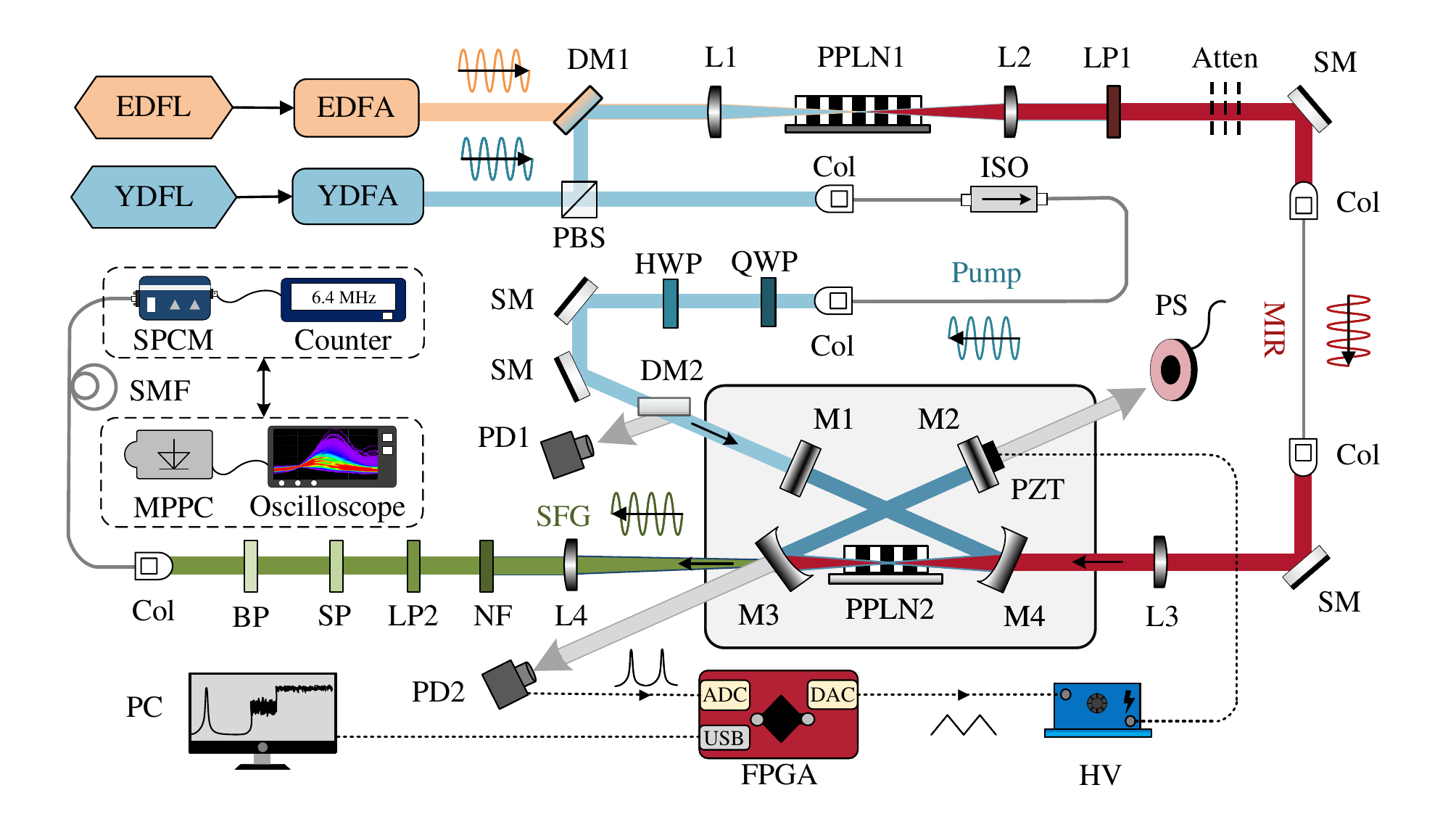}
\caption{Experimental setup of MIR upconversion detection based on the external-cavity pumping. An Er-doped fiber laser (EDFL) at 1550 nm and an Yb-doped fiber laser (YDFL) at 1064 nm are used to provide initial light sources for subsequent nonlinear frequency conversion. The two fiber lasers, operating at the single-longitudinal mode, are first used to prepare the MIR signal at 3.4 $\mu$m based on the difference-frequency generation in a periodically poled lithium niobate (PPLN1) crystal. Then the generated MIR beam is injected into an optical cavity for implementing the sum-frequency generation. The cavity comprises four mirrors and is stabilized with a digital locking unit based on a programmed field-programmable gate array (FPGA). Under the locked state, the pump power can significantly be enhanced within the cavity. After passing through a series of spectral filters, the upconverted signal is coupled into a single-mode fiber (SMF) before being detected by a single-photon counting module (SPCM) or a multi-pixel photon counter (MPPC). YDFA and EDFA, Yb- and Er-doped fiber amplifiers; DM: dichroic mirror; PBS: polarization beam splitter; L: lens; Atten, attenuator; SM, silver mirror; Col: collimator; ISO: isolator; HWP and QWP: half-wave and quarter-wave plates; M: cavity mirror; PZT: piezoelectric actuator; PD: photodiode; PC: computer; HV: High-voltage amplifier; PS: power sensor; NF: notch filter; LP, SP, and BP: long-pass, short-pass, and band-pass filters.}
\label{fig1}
\end{figure*}

Figure \ref{fig1} illustrates the experimental setup for the MIR upconversion detection system based on the external-cavity pumping enhancement. The light sources are from an Er-doped fiber laser (EDFL) at 1550 nm and an Yb-doped fiber laser (YDFL) at 1064 nm. The two fiber lasers are designed to stably operate at the single-frequency regime without suffering from the mode-hopping effect \cite{Fu2017JOSAB}. The linewidth of both EDFL and YDFL is less than 3 kHz. The EDFL and YDFL deliver CW linearly-polarized light at an average power of 24 and 27 mW, respectively. And the average powers are further boosted to 1 and 3 W by using Er-doped and Yb-doped fiber amplifiers (EDFL and YDFA). The output of YDFA is split by a polarization beam splitter. One portion with 1-W power is spatially combined with the amplified beam from EDFA by a dichroic mirror. The mixed beam is focused into a periodically poled lithium niobate (PPLN1) crystal with a length of 50 mm to perform the difference frequency generation (DFG). The operation temperature of the crystal is stabilized at 62 $^\circ$C with a precision of 0.1 $^\circ$C, which is optimized for the poling period of 30.49 $\mu$m. The generated MIR beam at 3.4 $\mu$m is collected into a single-mode fluoride fiber (Thorlabs, P3-32F-FC-1) with a coupling efficiency of 70\%. The average power for the generated CW MIR light is about 1 mW, and can be adjusted by selecting proper neutral density attenuators and varying the pumping power for the DFG \cite{Huang2021PR}. The precise MIR power calibration with a large dynamic range is essential for the subsequent characterization of the sensitive upconversion detection system. The prepared MIR source is focused into a bowtie-type standing-wave cavity by a calcium fluoride plano-convex lens (L3) with a focal length of 75 mm.

In parallel, the other potion from the YDFA serves as the pump source for the upconversion module. The pump laser is coupled into a single-mode fiber to improve spatial mode quality and beam-steering stability. A fiber optical isolator is used to prevent returning light from disturbing or damaging the fiber laser. The pump beam is then collimated via an aspheric lens with a 7.5-mm focal length after an FC/APC fiber patch cord. The polarization state of the pump beam can be adjusted by rotating the half-wave and quarter-wave plates. The pump beam is mode-matched to the external cavity by optimizing the separation between the fiber end and the lens, as well as the distance between the collimator and the cavity. To perform the sum frequency generation (SFG), another nonlinear crystal (PPLN2) is used with a length of 25 mm and a poling period of 22.4 $\mu$m. The upconverted SFG light at 810 nm is steered through a spectral filtering stage that consists of a notch filter at 1064 nm with a bandwidth of 44 nm, a long-pass filter with a cut-off wavelength of 700 nm, a short-pass filter at 900 nm, and a band-pass filter with a bandwidth of 3 nm. The total transmission efficiency $\eta_\text{filter}$ of the filtering group is about 64.8\% with a rejection ratio for the pump about 190 dB. Finally, the filtered signal is coupled into a single-mode fiber with coupling efficiency $\eta_\text{fiber}$ of 80\% before being recorded by a single-photon counting module (SPCM) based on silicon avalanche photodiode (APD) with a detection efficiency $\eta_\text{det}$ of 60\%.

\begin{figure*}[t!]
\includegraphics[width=0.75\textwidth]{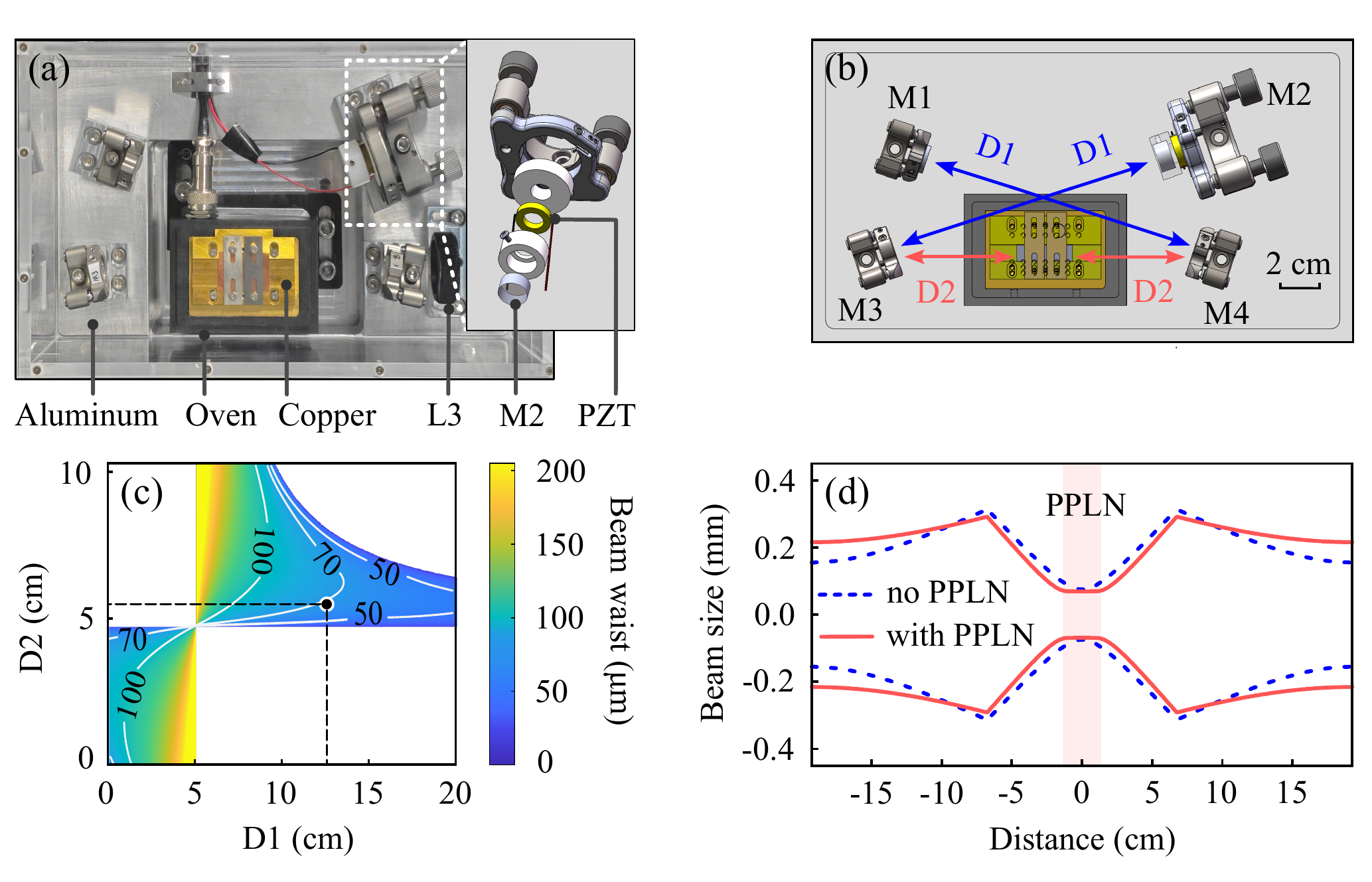}
\caption{(a) Photo of the enhancement optical cavity, indicating the materials of each part. (b) Physical layout of the bowtie cavity. The distances D1 and D2 are two critical parameters for the cavity design. (c) The radius of the beam waist within the nonlinear crystal as a function of D1 and D2. The black point denotes the values used in the experiment, corresponding to a beam radius of 69 $\mu$m. Note that the area in white represents the unstable region for an optical cavity. (d) Evolution of the beam size as the light propagates along the cavity, with and without the presence of the crystal inside the cavity. The origin is defined at the center of the crystal. The shaded area indicates the occupied space by the 25-mm-length nonlinear crystal.}
\label{fig2}
\end{figure*}

\section{Results and discussion}
In the following, we begin with characterizing the optical cavity. Figure \ref{fig2}(a) presents the photograph of the implemented cavity, which consists of four mirrors on an aluminum base. Specifically, an oven is inserted within the cavity, where the crystal is placed on the copper holder. The oven is mounted on a high-precision alignment stage (Newport, 9071-M), which allows a four-axis adjustment with a 3-mm linear movement and an 8$^\circ$ angular rotation. The compact kinematic stage facilitates the alignment of the nonlinear crystal in a precise and stable way. Notably, one mirror (M2) is mounted with a donut-shaped piezoelectric actuator (PZT, Thorlabs PA44M3KW), which gives an auxiliary port for monitoring the leaking pump. As depicted in Fig. \ref{fig2}(b), the resonator is integrated into an assembly of $24\times 13\times 11$ cm$^3$. Mirrors M1-3 and M4 are made of N-BK7 and calcium fluoride, respectively. The M1 and M2 are flat mirrors, and M3 and M4 have a radius of curvature of 100 mm. The coupling mirror M1 is coated with a power reflection of 97\% at 1064 nm, while the others are highly reflected. More details about mirror design are given in Table \ref{tab1}.

\begin{table*}[t!]
\renewcommand\arraystretch{2}
\setlength{\tabcolsep}{20pt}
\caption{Parameters of the cavity mirrors.}
\label{tab1}
\centering
\begin{tabular*}{0.8\linewidth}{@{}ccccc@{}}
    \hline                     
    \rule{0pt}{10pt} 	
	Mirror & Material & ROC$^{[a]}$ (mm) & S1$^{[b]}$ & S2$^{[b]}$\\
	\hline
	1 & N-BK7 & $\infty$ & PR($p$) & AR($p,s,u$)\\
	2 & N-BK7 & $\infty$ & HR($p$) & AR($p,s,u$)\\
	3 & N-BK7 & 100 & HR($p$), HT($u$) & AR($p,s,u$)\\
	4 & CaF$_{2}$ & 100 & HR($p$), HT($s$) & AR($p,s,u$)\\
	\hline
    \end{tabular*}
    \begin{tablenotes}
\item[] $^{[a]}$ROC: The radius of curvature of the cavity mirror.
\item[] $^{[b]}$S1 and S2 are the two surfaces of the cavity mirror. HR (high reflection), AR (anti-reflection), and PR (partial reflection) designations are followed by the wavelength for which the coating was designed, using the following abbreviations: pump ($p$, 1064 nm), signal ($s$, 3400 nm), upconverted ($u$, 810 nm) fields.
\end{tablenotes}
\end{table*}

In the cavity configuration, the mirror separations are crucial for adjusting the cavity-mode size, as shown in Fig. \ref{fig2}(c). In our experiment, the D1 and D2 are set to be 12.5 and 5.5 cm, which results in a cavity waist of 69 $\mu$m. The tight waist is beneficial for achieving high pump intensity, while the corresponding Rayleigh length for the Gaussian mode is sufficiently long to avoid the crystal insertion loss. Figure \ref{fig2}(d) presents the beam-size evolution along the optical cavity. For a crystal thickness of 1 mm, the beam can propagate through the nonlinear crystal without suffering from beam truncation.

\begin{figure*}[t!]
\includegraphics[width=0.85 \textwidth]{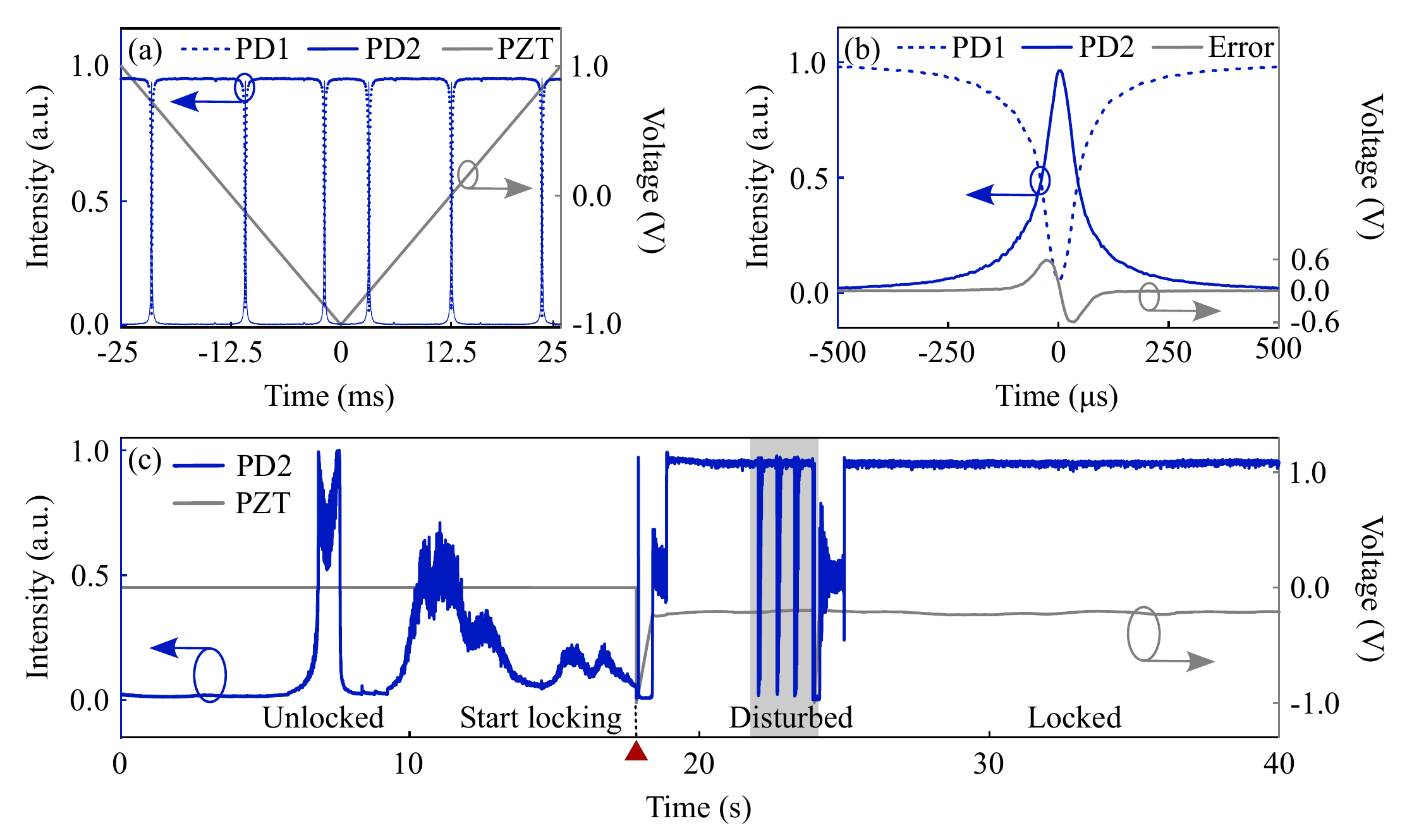}
\caption{(a) Recorded traces of the reflection (PD1) and transmission (PD2) light from the optical cavity, as scanning the PZT at a rate of 20 Hz. The peaks correspond to the resonating points for the fundamental spatial mode. (b) Enlarged illustration for the peak, along with the error signal for the dither locking. (c) Measured intensity for the cavity transmission during the stabilization process. The locking is engaged at 18 s, as indicated by the red triangle. The presented disturbing drops are ascribed to the intentional perturbation for testing the re-locking feature of the digital feedback unit.}
\label{fig3}
\end{figure*}

To further analyze the performance of power enhancement, a high-voltage sawtooth waveform is applied to the PZT for cavity-length sweeping. As given in Fig. \ref{fig3}(a), a series of resonating peaks can be observed for the pump transmission (PD2) and reflection (PD1). The uneven separation between two adjacent peaks is due to the nonlinearity of the PZT actuator. Moreover, the hysteresis behavior for the PZT is manifested by the asymmetry for the up- and down-scanning traces. Thanks to the spatial mode filtering for the pump, a nearly perfect mode matching is achieved with negligible high-order peaks. The presence of only the fundamental  mode would facilitate the cavity locking at a stable transverse mode. The cavity finesse $\mathcal{F}$ is evaluated to be about 109 from the ratio between the free spectral range and the cavity bandwidth. Note that average values for the peak separation and peak width are used to improve the estimation accuracy. The finesse is directly related to the cavity loss, as given by the following formula \cite{Witinski2009AO}:
\begin{equation}
\mathcal{F} =\frac{2\pi}{T_{1}+T_{2}+2T_{3}+2T_{4}+2L} \ ,
\end{equation}
where $T_i$ is the power transmission for each mirror, and $L$ is the insertion loss for the crystal. The expression for peak intensity enhancement is given by \cite{Witinski2009AO}:
\begin{equation}
	\Gamma=\frac{1-R_1}{(1-\sqrt{R_1 R_2} R_3 R_4 T)^2} \ ,
\end{equation}
where $R_i$ is the power reflectivity for each mirror, and $T$ is the transmission for the crystal. At the approximation of a low-loss cavity, the enhancing factor is directly linked to the cavity finesse as $\Gamma  \approx \mathcal{F}/\pi$. In our experiment, the enhancing factor is measured to be about 36, which is consistent with the theoretical value of 35.6. Although the standing-wave cavity is more lossy than the traveling-wave cavity, the bidirectional operation offers a better immunity to the perturbation of fractional Bragg reflection from a periodically-poled nonlinear crystal \cite{Huang2019OE}. The standing-wave cavity layout could provide an expedient solution to obtain more stable and higher enhancement, especially at the presence of long-length crystals. Additionally, the enhancement cavity is engineered to be impedance matched as shown by the reflection trace (PD1) in Fig. \ref{fig3}(b), where the transmission coefficient of the input mirror (M1) for the pump radiation matches the coefficient quantifying all other losses. Consequently, the back reflection of pump power is minimized to improve the energy utilization efficiency and to reduce the detrimental effect on the amplifier and laser.

In our experiment, the active stabilization of the external cavity is realized by a digital dither locking system based on an FPGA hardware board \cite{Neuhaus2017EQEC}. The FPGA is programmed to output a small-amplitude sinusoidal waveform with a frequency of 20 kHz. The dithering operation on the PZT behaves as if the laser frequency were oscillating back and forth. A digital lock-in amplifier is used to produce an error signal, as shown in Fig. \ref{fig3}(b). Such a digital locking system offers unique features of state monitoring and automatic operation. Figure \ref{fig3}(c) illustrates a representative process from unlocked to locked states. The several disturbing drops are ascribed to the intentional perturbation, which indicates a fast re-locking capability. When the cavity is locked at the resonance for an injection pump power of 1.53 W, the circulating power in the cavity is inferred to be about 55 W from the pump leakage out of the cavity mirror (M2). The power fluctuation of the enhanced pump light is estimated to be about 1\%, which is essential to implement a high-stability frequency upconverter.

\begin{figure*}[t!]
\includegraphics[width=0.75 \textwidth]{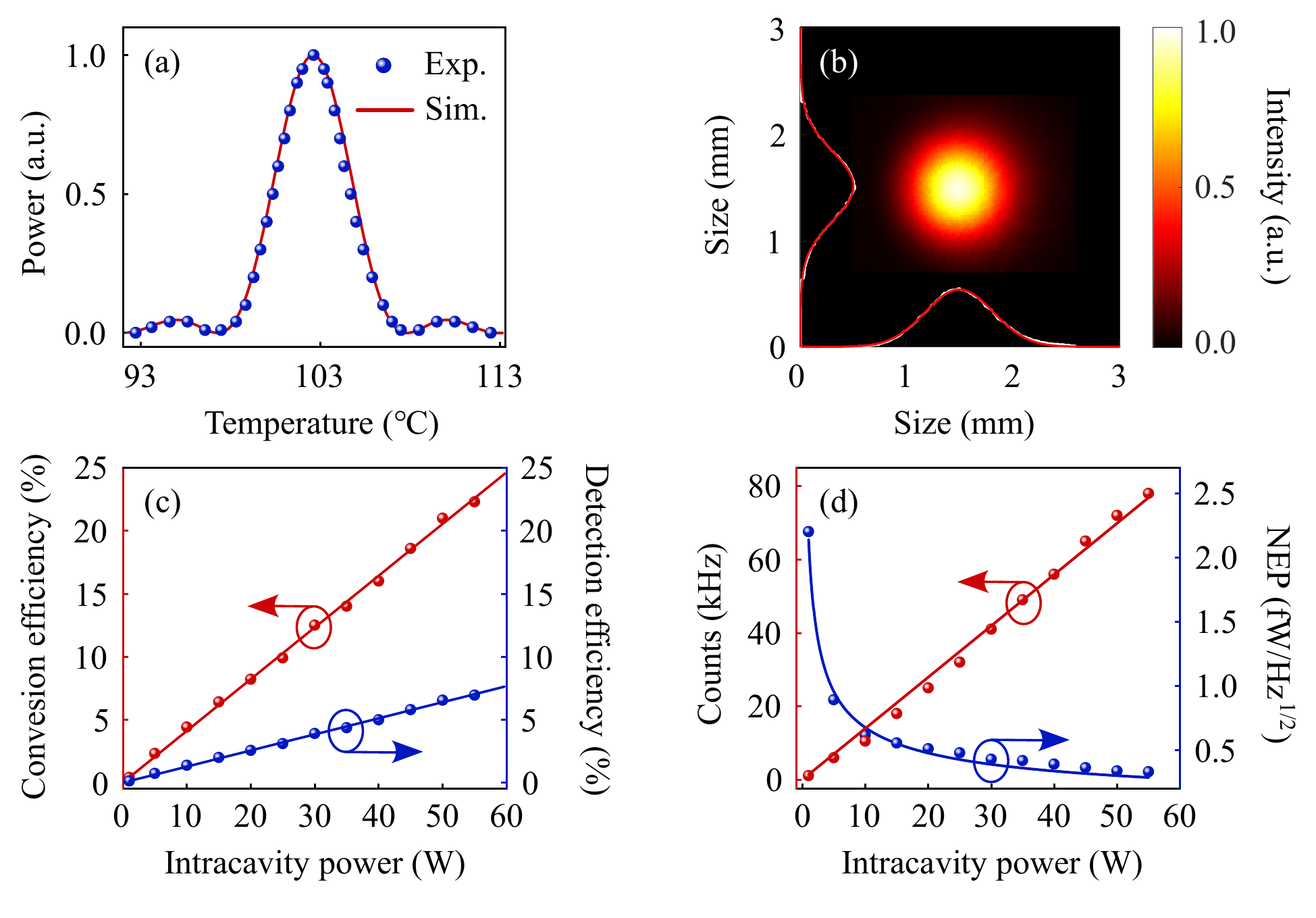}
\caption{(a) SFG power as a function of crystal temperature. (b) Captured beam profile for the SFG light. (c) Intrinsic conversion efficiency and total detection efficiency increase as augmenting the intracavity power. (d) Measured background count and equivalent noise power as a function of the intracavity power.}
\label{fig4}
\end{figure*}

\begin{table*}[t]
\renewcommand\arraystretch{2}
\setlength{\tabcolsep}{13.5pt}
\caption{Performance comparison for CW-wave pumping MIR upconversion detection systems.}
\label{tab2}
\begin{tabular*}{0.85\linewidth}{@{}ccccc@{}}
\hline
 Ref. & Scheme & Wavelength ($\mu$m) & NEP (fW/Hz$^{1/2}$) & Conversion efficiency (\%)  \\ 
 \hline
 This & External-cavity & 3.4 & 0.3 & 22 \\ 
\cite{Wolf2017OE} & External-cavity & 3.7-4.7 & 700-10000 & $ \sim $ 0.001\\
\cite{Li2024OLT} & Single-pass & 4.6 & 3.24 & 8.87\\ 
\cite{Ge2023PRA} & Single-pass & 4.15 & \quad /$^{[a]}$ & 0.18 \\ 
\cite{Demur2017OL} & Single-pass & 10.6 & 4000 & 20 \\  
 \cite{Dam2012NP} & Intra-cavity & 3 & \quad /$^{[b]}$ & 20 \\
\cite{Pedersen2019PTL} & Intra-cavity & 3.4 & 20  & 6\\
\cite{Huang2017LP} & Intra-cavity& 3 & \quad /$^{[c]}$ & 28$^{[d]}$  \\
\cite{Widarsson2022AO} & Intra-cavity& 3 & 161$^{[e]}$ & 2 \\
\hline
\end{tabular*}
\begin{tablenotes}
\item[] $^{[a]}$The minimum detectable incoherent MIR power of the system is 813 fW, but the NEP is not specified.
\item[] $^{[b]}$The dark noise is 0.2 photons/spacial element/second when operating at room temperature, and the used camera has 200$\times$100 resolvable image elements. The system detection efficiency is not specified.
\item[] $^{[c]}$The minimum detectable incoherent MIR power of the system is 31 fW, but the NEP is unspecified.
\item[] $^{[d]}$The conversion efficiency per watt of pump power is comparable to our work.
\item[] $^{[e]}$The NEP is inferred from the given values with a background noise of 8.5 MHz and a system detection efficiency of 0.15\%.
\end{tablenotes}
\end{table*}

Now, we turn to investigate the performance of the MIR upconversion detector. Figure \ref{fig4}(a) shows the measured dependence of the SFG intensity on the operating temperature of the nonlinear crystal, which agrees with the theoretical simulation under the assumption of plane-wave interaction. With the 22.4-$\mu$m grating period of the PPLN crystal, the phase-matching temperature is identified to be 102.6 $^\circ$C. Figure \ref{fig4}(b) shows the beam profile of the upconverted light when the optical cavity is locked at the resonance to the pump light. The cross sections along the two orthogonal directions indicate a superior Gaussian spatial mode confined by the optical resonator. The high beam quality of the pump field not only improves the conversion efficiency due to the improved spatial overlap among interacting fields, but also facilitates subsequent coupling into a single-mode fiber for suppressing background noises. Under the approximation of non-depleted pumping, the conversion efficiency is defined by the photon-flux ratio between the SFG and MIR light, which is expressed as
\begin{equation}
	\eta_\text{conv}=\frac{P_u}{P_s} \times \frac{\lambda_u}{\lambda_s} \ ,
\end{equation}
where $\lambda_u$ and $\lambda_s$ are the wavelengths for SFG and MIR light, and $P_u$ and $P_s$ denote the measured average power for the two fields. As shown in Fig. \ref{fig4}(c), the conversion efficiency is increased with a higher intracavity power. Correspondingly, the peak total detection efficiency for the MIR signal is evaluated to about 6.8\% according to $\eta_\text{total} = \eta_\text{conv} \times \eta_\text{filter} \times \eta_\text{fiber} \times \eta_\text{det}$.

Another key parameter for an optical detector is the background noise, which is measured by blocking the MIR signal before the upconversion detector. As shown in Fig. \ref{fig4}(d), the background noise linearly increases as augmenting the pump power. The photon count of the noise $N_\text{BG}$ is measured to 78 kHz at an intracavity power of 55 W. The background noise is mainly due to the upconverted light of the spontaneous parametric down-conversion fluorescence \cite{Pedersen2019PTL}. For a photon-counting detector, the detection sensitivity is usually defined by the noise equivalent power (NEP) \cite{Pedersen2019PTL, Huang2021PR}:
\begin{equation} 
\sigma_\text{NEP} = \frac{E\sqrt{2N_\text{BG}}}{\eta_\text{total}} \ ,
\end{equation}
where $E$ represents the photon energy of the MIR signal. As shown in Fig. \ref{fig4}(d), the $\sigma_\text{NEP}$ tends to saturate for high pump powers. The achieved minimum NEP is about 0.3 fW/Hz$^{1/2}$, which is at least one order of magnitude better than reported values in various configurations as listed in Table \ref{tab2}. It can seen that our scheme not only provides a high conversion efficiency due to the significant pump enhancement, but also exhibits a high detection sensitivity due to the effective noise suppression. Compared to the intra-cavity scheme, the presented approach can in principle support a higher enhancing factor due to much fewer elements in the cavity. With the help of further loss reduction, the intracavity pump power up to 400 W is feasible with an enhancing factor up to 100 \cite{Wolf2017OE}. Another unique feature of the external-cavity enhancement is the exclusion of the laser pump at a wavelength close to the SFG signal, which hence leads to a much lower background noise. We note that the upconverted thermal emission also contributes  to the background noise, which is particularly pronounced for operation wavelengths above 3 $\mu$m due to the increased linear absorption for the PPLN crystal \cite{Pedersen2019PTL}. The thermal noise can be suppressed by reducing the crystal temperature at an optimized poling period for the phase matching.

\begin{figure}[b!]
\includegraphics[width=0.85 \columnwidth]{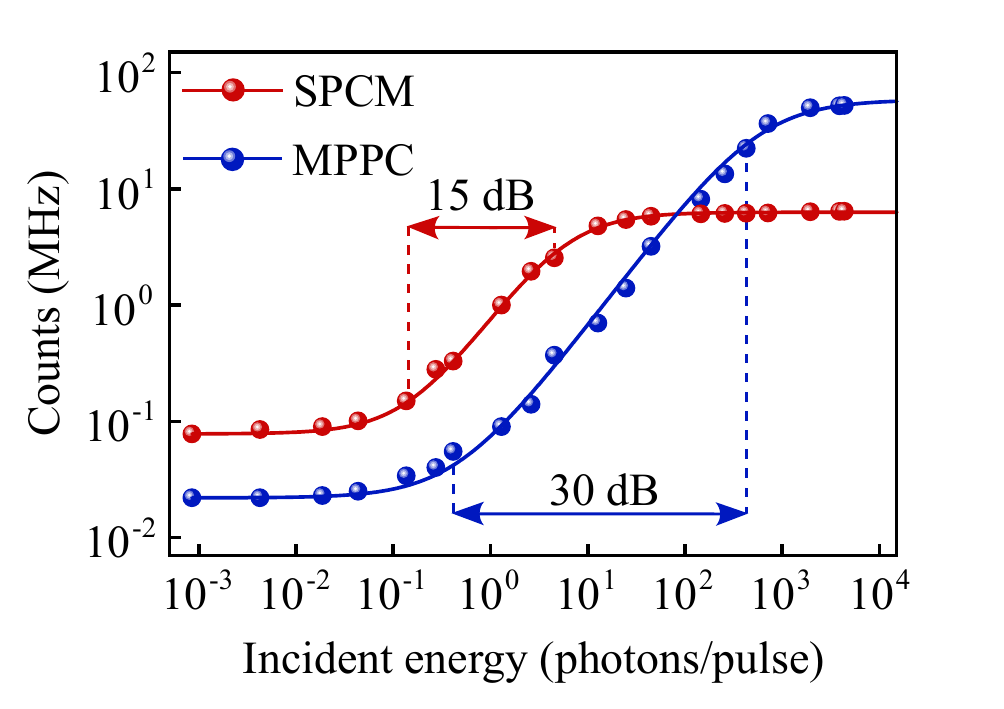}
\caption{Measured photon counts as increasing the incident power in two cases of using optical detectors based on SPCM and MPPC, which indicate detection dynamic ranges of 15 dB and 30 dB in the linear response regime, respectively. The solid lines are fitted to guide the eyes.}
\label{fig5}
\end{figure}

Finally, we have adopted a silicon MPPC for extending the detection dynamic range for the MIR upconversion detector. In contrast to the single-pixel photon counter, the MPPC offers a unique capability of photon-number-resolving (PNR) detection \cite{Huang2021PR}. The MPPC (Hamamatsu Photonics, S13362-3050DG) comprises 60$\times$60 pixels on an effective active area of 9 mm$^2$. The PNR detection is realized via spatial multiplexing, where incident photons could be registered by different pixels. The detection efficiency $\eta_\text{MPPC}$ MPPC was about 7\% at 810 nm, which can be further improved by resorting to an MPPC with a larger pixel pitch and a higher fill factor. To better characterize the detector performance, a pulsed MIR source is prepared by inserting an electro-optic modulator (EOM, not shown in Fig. \ref{fig1}) at the output of EDFL, which generates a pulse duration of 1 ns at a repetition rate of 10 MHz. As shown in Fig. \ref{fig5}, the photon-counting rate for the SPCM is saturated as increasing the incident power, which results in a dynamic range of 15 dB. In comparison, the MPPC offers a dynamic range of up to 30 dB. Notably, the MPPC typically has a large detection area, which favors a high collection efficiency for the upconverted signal in the free-space configuration. The achieved MIR detection performance with single-photon sensitivity and large dynamic range would facilitate a variety of applications, such as free-space communication, trace molecule spectroscopy, and long-range infrared sensing.

\section{Conclusion}
In conclusion, we have implemented an ultra-sensitive MIR upconversion detection system with high conversion efficiency and low background noise. The operation wavelength can cover a MIR spectral range from 3-5 $\mu$m defined by the transparency window of the PPLN crystal. The demonstrated sensitivity is at least two orders of magnitude better than conventional detectors based on PbS, PbSe, and HgCdTe \cite{Russo2022Photonics, Wang2019Small}. In comparison to single-pass schemes \cite{Demur2017OL, Li2024OLT, Ge2023PRA}, the involved nonlinear converter is constructed based on the external optical cavity, which favors pump-power enhancement, spatial-mode confinement and parametric noise suppression. Moreover, the presented configuration favors a reduced cavity loss due to fewer optical elements, which can in principle access high enhancing factors that are beyond the reach of the intra-cavity scheme \cite{Widarsson2022AO, Pedersen2019PTL}. Notably, the separation arrangement of the laser and enhancement cavities allows one to adopt the compact single-frequency fiber lasers and amplifiers \cite{Fu2017JOSAB}, which circumvents the complexity of spectral engineering in the solid-state lasers. Such a narrow-band pumping is essential for achieving high-fidelity spectral mapping, which is particularly attractive for high-precision molecular spectroscopy \cite{Neely2012OL}. In this scenario, our MIR upconversion detector can be modified to obtain a large spectral coverage by using nonlinear crystals with fanout \cite{Barh2019OL} or chirped poling \cite{Friis2019OL, Mrejen2020LPR} structures. In addition, the operation window is possible to extend into longer infrared wavelengths up to 12 $\mu$m with BaGa$_4$Se$_7$ \cite{Liu2022Optica} or AgGaS$_2$ \cite{Rodrigo2021LPR} crystals. It is worth noting that the presented architecture is readily adapted to realize upconversion MIR imaging \cite{Huang2022NC, Dam2012NP}.

\vspace{8pt}
\noindent  {\fontfamily{phv}\selectfont 
\normalsize \textbf{Acknowledgments.} 
}
\noindent This work was supported by National Key Research and Development Program (2021YFB2801100), National Natural Science Foundation of China (62175064, 62235019, 62035005, 12022411); Shanghai Pilot Program for Basic Research (TQ20220104); Natural Science Foundation of Chongqing (CSTB2023NSCQ-JQX0011, CSTB2022NSCQ-MSX0451, CSTB2022NSCQ-JQX0016); Shanghai Municipal Science and Technology Major Project (2019SHZDZX01); Fundamental Research Funds for the Central Universities.

\vspace{8pt}
\noindent  {\fontfamily{phv}\selectfont 
\normalsize \textbf{Disclosures.} 
}
\noindent The authors declare no competing interests.

\vspace{8pt}
\noindent  {\fontfamily{phv}\selectfont 
\normalsize \textbf{Data availability.} 
}
\noindent The data that support the findings of this study are available from the corresponding author upon reasonable request.

\end{document}